\documentclass[titlepage,twoside,letterpaper,12pt]{article}
\usepackage{fancyhdr}
\usepackage{color}
\usepackage{url}
\usepackage{graphicx}
\usepackage{caption}
\begin{document}

\begin{titlepage}
\centering
\vskip 0.1in

{\Large{\bf{Indoor Information Retrieval using Lifelog Data}}}

\vskip 0.5in
{\large{Deepanwita Datta}} \\
{\em{Indian Institute of Technology (BHU), Varanasi, India}}  
\vskip 0.2in

\vskip 0.2in
\large{September 10, 2018}
\end{titlepage}
\section{Summary}
Studying human behaviour through lifelogging has seen an increase in attention from researchers over the past decade. The opportunities that lifelogging offers are based on the fact that a lifelog, as a ``black box'' of our lives, offers rich contextual information, which has been an Achilles heel of information discovery. While lifelog data has been put to use in various contexts, its application to indoor environment scenario remains unexplored. In this proposal, I plan to design a method that enables us to capture and record indoor lifelog data of a person's life in order to facilitate healthcare systems, emergency response, item tracking \emph{etc}. To this end, we aim to build an Indoor Information Retrieval system that can be queried with natural language queries over lifelog data. Judicious use of the lifelog data for the indoor application may enable us to solve very fundamental but non-avoidable problems of our daily life. Analysis of lifelog data coupled with Information Retrieval is not only a promising research topic, but the possibility of its indoor application especially for healthcare, lost-item tracking would be an innovative research idea to the best of our knowledge.
\newpage
\section{Introduction}
The rapid growth of personal devices, such as smartphones, video cameras as well as different wearable devices makes available a massive volume of personal data, i.e., Lifelog data, by capturing pictures, videos, and audio clips in every moment of our life. Certainly, there arises a need for an automatic system which can analyze this enormous amount of data in order to categorize, summarize and facilitate querying to retrieve the information the user may need. Such a system may aid people to retrieve a specific moment/memory of interest by analyzing their lifelog data and processing queries on lifelog data. These queries can be of various types and modalities. For the sake of our research we choose to concentrate on indoor lifelog data, on top of which we then want to build an Information Retrieval system that will help user in answering personalized queries based on past events. 
 
 However, for devising such an application, we may need to focus on some related on-going research problems and come up with some efficient and suitable solutions. To realize our objective, we need to study the problem closely by segmenting into individual but related research questions:
\begin{itemize}
    \item{\textbf{Lifelog Data Gathering and Processing:}}
 The indoor activities of a person that is captured via video sequences, frames or images would be the first step towards data gathering. This data may very well be augmented with biometrics coming from various sensors, such as heart rate, galvanic skin response, calorie burn, steps.
Further, it would be interesting to obtain the person’s on-screen activity logs like music listening history, computer usage, the frequency of typed words via the keyboard and information consumed on the computer, \emph{etc}. as this could provide valuable information about the person’s behaviour. The reason for collecting biometric data is to precisely determine the last event in which the item to track was involved. Say, a carpenter, who regularly uses his hammer suddenly loses it. While we may, if available, backtrack the video footage of the person last using the hammer, it is certainly cumbersome. It is evident that each intense activity is associated with certain biometric characteristics such as an increase in heart rate while doing an intense physical activity \cite{Kafri2014}. So, if we can trace back to the last time such a biometric signature was produced, we can begin tracing the item from that timestamp onwards.
 
Since the data collected is enormous and sifting through the data would entail both time and effort, it is essential to identify “concepts” and categorize them. By concept, we are referring to the actions and events related to the item or activity in question. For example, if the missing item is a wallet, all and every event that involves the wallet for a certain past period (timeframe) would constitute a concept. In that particular scenario, all the other data is irrelevant. Each item leaves a trail of concepts which need to be assembled and grouped. This calls for classification and categorization of video, audio or image sequences along with the supplementary data. Different computer vision techniques may be applied to detect locations, actions and to recognize different entities from visual data.

\item{\textbf{Cross-Modal Retrieval:}}  Cross-modal retrieval has proven to be an effective solution while searching over multi-varied data especially where the data size is enormous \cite{Cavallo2018}. A typical example of cross-media retrieval is when an image is searched through a text query. The retrieval system should be able to judiciously determine which modalities (from lifelogging, activity and biometric data) to use for a certain item and combine them to enable an efficient search. Each modality in itself comprises certain aspects or features. These can broadly be classified as:
\begin{itemize}
    \item{(a.)} Temporal aspects like the date and time, time span or period, frequency, \emph{etc}.  \item{(b.)} Scene aspects like scene category, scene attributes
\item{(c.)} Entity and Action aspects like the existence of entities, activities of the user and 
\item{(d.)} Augmented aspects like biometric data, computer usage information, \emph{etc}.
\end{itemize}
Any lifelogging retrieval system that adopts cross-modal retrieval as the lifelog data which is captured by various sensors, is inherently multimodal by nature. For such a lifelogging retrieval system, the steams of multimodal lifelog data is necessary needed to be aligned and synchronized. Also, the multimodal data should be combined into some form of retrievable units. Thus the first and most important step of lifelogging data retrieval is to come up with an indexable unit of individual’s activity captured by various sensors, as proper indexing is the key of any retrieval system both in terms of efficiency and time. (Such an indexable unit would manifest as a fusion of a number of synchronized lifelog data sources at a given point in time, such as audio, video or images from wearable cameras, readings from biometric sensors, communication records, information creation or access logs, and various other sensor sources. Such a unit would represent a contiguous period of time.) Among all theses multimodal data, textual documents is most easier to process and convenient mode for retrieval. So if we can generate or extract some textual description/document from the continuous streams of lifelog data coming oﬀ various sensors and wearable devices, the formation of indexable units will be much easier.\\
Now the questions arrive like :
\begin{itemize}
    \item{(1.)} What a lifelog textual description or document should be?
    \item{(2.)} How to generate such document units from the continuous streams of lifelog data?
\end{itemize}
Lately, a lot of research is going on to answer these above questions. Also, we have come up with a few ideas which we have discussed in detail in the section \ref{WP}.

\item{\textbf{Multimodal Query Reformulation:}}
Assuming that the query is provided as a single modality, it may be imperative to convert it to or add other modalities to the search query to make it closer to the user's information need. Although a significant amount of work has focused on bridging the semantic gap between the high-level information need of users and commonly employed low-level features, it continues to be an open challenge.
Correlating different sources of information e.g. textual and visual information is not a trivial task. The available state-of-the-art addressing this challenge is two-pronged. Few of these existing works pay attention to learning mapping functions among the different modalities of data whereas the rest of them explore the high-level semantic representation of modalities. Among these semantic representation based approaches, deep learning based approaches have gained popularity in recent times. Generally, deep convolutional neural networks are used to learn the latent features, and these learnt features are utilized as semantic representation in these models. Once the proper representation is obtained, a multimodal query can be formulated from different extracted features.
 
\item{\textbf{Cross-Lingual Information Retrieval:}} For making the system robust, Cross-Lingual retrieval can be incorporated. This could be an enhancement once the proposed system is in place. In real-life users provide queries in textual form which is the most accessible and most prevalent form of input. Confining the system's working to a particular language would not be beneficial if we are to aim for wide reachability and usability. Hence the system should be inclusive of various (major) languages. Thus the idea of Cross-Lingual Information retrieval. While the problem is a long-standing one and has been researched upon widely our focus would be on Translated text-query based retrieval vs. Native Language Query based retrieval performance. Intuitively, the former should yield better performance than the latter, in most cases. The challenge would be to enhance the Native Language query-based information retrieval performance to make it at par with the former or better yet to exceed it. The challenge arises because most translators take the literal sense of the terms in the query whereas a native language query can semantically interpret the information need. Sometimes, this problem is compounded because a particular word may not have a proper English language translation or vice-versa. This task would invariably involve some Natural Language Processing.
 
\item{\textbf{Application(s):}} Possible real-life applications of the system could very well be extended to outdoor situations such as tracking missing luggage at airports or railway stations. Also, in recent days, an exponential growth of stress and busy lifestyle are causing several neurological diseases like Dementia. There is an increase in the number of patients suffering from Alzheimer’s disease which causes severe memory loss. So, it would be a great help if we could assist those patients to find their lost items on a daily basis by capturing and analyzing their lifelog. Another possible application could be proactively contacting emergency services on behalf of patients who have a history of, say, seizures or heart attacks, in case there are symptoms (indicators from recorded data) of the same. Of course that would need a real-time response system, which would be our focus once the standard system is in place.

\end{itemize}

\section{Background}
Lifelogging represents a phenomenon whereby people can digitally record their own daily lives in varying amounts of detail, for a variety of purposes. Capturing a human's life activities, mining and inferring knowledge from them provides a comprehensive understanding of how one leads and lives a life \cite{Gurrin2014}. The benefits of lifelogging can range from content-based information retrieval, contextual retrieval, browsing, search, linking, summarisation \emph{etc.}, user interaction to acting as a memory aid \cite{Bahrainian2018}. However, there are challenges in managing, analyzing, indexing and providing content-based access to streams of multimodal information derived from lifelog sensors which can be noisy, error-prone and with gaps in continuity due to sensor calibration or failure.
If we know a detailed context of the user (for example, who the user is, where she is and has been recently, what she is doing now and has done, who she is
with, \emph{etc}.) then we could leverage this context to develop more useful
tools for information access.
The rapid growth of inexpensive sensing technologies and availability of significantly cheaper computer storage have fostered lifelogging to become a mainstream research topic. Although the recent advancements in sensing technology empower the efficient sensing of personal activities, locations, environment \emph{etc.}, lifelogging imposes an open challenge to the information retrieval community. The primary challenge lies to manage, analyze, index and to provide the content-based access to the streams of multimodal information derived from different lifelog sensors. Apart from that, lifelogging also suffers from lifelog sensors which often produce noisy data and they are also error-prone. Additionally there lies various sensory gaps between the sensor and the interpreter. Besides these drawbacks lifelogging offers benefits to retrieve information, browse the web, summaries information and to handle the user interaction. Moreover, judicious use of lifelogging data may aid the content-based information retrieval and contextual retrieval as lifelogging data offers rich contextual information. For example, if we know a detailed context of the user, like, who the user is, where she is and has been recently, what she is doing now and has done, who she is with, \emph{etc.} then we could leverage this context to develop more useful tools for information access. Exploiting this valuable contextual information provided by lifelogging to the field of information retrieval has received little research attention to date.\\
We have been using the term `lifelogging' frequently for quite a while since the very beginning of our proposal. For a better understanding, we briefly explain some of the common terminologies used in this proposal. \emph{Lifelogging} is the process to gather the data from life activity/experience of an individual by a variety of sensors and to process them where a \emph{Lifelog} is the actual data gathered by a lifelogger. Another closely associated term is \emph{Surrogate Memory} which is a simple digital library to store the lifelog data. It also includes some software to organize and manage a list of events or episodes from the lifelog. While talking regarding lifelogging data collection, one key point is to be noted that the collection process typically should be taken passively, no intervention of lifelogger should be needed to initiate anything. Although nowadays a few dedicated individuals are willing to log their life, their number is still insignificant. Another problem of such data collection is, in such cases sometimes people explicitly captures some moments where they become conscious. Such data does not reflect the natural activities of any individual, like when we pose smile at some party that does not reflect our actual mental state at that moment. So, lifelogging must also contain those data which are captured involuntarily or rather captured passively to avoid the conflict between conscious and habitual activity of an individual. Thus lifelog contains such inadvertent activities which include the records of everything an individual has done, are mostly repetitive and increase the volume of lifelog data. \\
In effect, this research domain has attracted serious traction over the last few years and its applications, as stated earlier, are myriad. One important aspect of sifting through lifelog data especially those arising from visual sensors is to segment them into separate events. Doherty and Smeaton \cite{Doherty2008} improved the deconstruction of a substantial collection of images (captured using SenseCam, a personal wearable device) to discrete events. They did so through not only the introduction of some intelligent threshold selection techniques, but also through improvements in the selection of normalization, fusion, and vector distance techniques. We intend to build upon this idea since if we are to design an IR system that helps in tracking missing items, we would essentially need to review that person's last $n$ events or actions.
With the advent of the smartphone era, applications on smartphones can capture various sensor data which can then be used for several applications such as location tracking \cite{Kalnikaite2010}, health monitoring \cite{Katz2018}\cite{Yang2018}, child development monitoring \cite{Kientz2009}, \emph{etc}. This availability of cheap and accessible data has inherently opened avenues for interdisciplinary research from varied domains such as cognitive reasoning, computational social science, ubiquitous computing, public health and so on. However, as with any technology, lifelogging too suffers from inadequate uptake primarily because of lack of a uniform or standardized design principle. Whittaker \emph{et al.}~\cite{Whittaker2012} lay down few such principles after a rigorous study namely Selectivity, Embodiment, Synergy, and Reminiscence. The authors show that design principles were generative, leading to the development of new classes of lifelogging system, as well as providing strategic guidance about how those systems should be built. Searching and retrieving lifelog data has also been studied in various contexts such as video retrieval \cite{Hori2003}, recorded audio retrieval \cite{Shah2012}, image retrieval \cite{Doherty2008b} and so on. Recently a test collection for lifelog research was published by NTCIR that is designed to index and retrieve multimodal lifelog data \cite{Gurrin2016}.
\section{Experimental requirements}
In this section, we list the possible requirements in terms of hardware and software. The specifications are only indicative and may increase or decrease as per the requirement of the project. There is a wide range of existing devices, either portable or stand devices, that can be used for lifelogging. A short list of such available devices is noted down below: 
\begin{itemize}
    \item{Wearable Cameras:} Such wearable cameras capture photos of what the individuals see in front of them. The mode of photo capturing is mostly consequent and passive. \emph{Ex.} SenseCam, Video glass, Looxcie, Go-Pro, head-mounted camera \emph{etc.} 
    \item{Biometric Devices:} Biometric devices are mostly sensor based. Human body conditions are sensed and collected from galvanic skin response (GSR) and skin temperature (ST), from different physiological responses such as heart rate or increased sweat production, sympathetic nervous activity \emph{etc.} \cite{kang11}.
    A few examples of such devices are Polar Heart Rate Monitor, ReadiBand, BodyMedia SenseWear Armband \emph{etc.}. These devices monitor heart-rate and sleeping patterns from wrist. Also devices like Foster-Miller vests track respiration, body temperature, heart rate, gps \emph{etc}. Another device, Posture monitoring vest, measures body bending on structural beams by using several wearable plastic optical fiber sensor outside the garment.
    \item{Fitness Devices:} Fitness devices are very popular these days. Fit-Bit, Nike+Pod \emph{etc.} are the examples of such small and comfortable devices. They can help us record various health-related metrics such as metabolic rate, pulse and so on.
    \item{Non-visual Wearable Devices:} Such devices are non-visual because they do not take any visual content about individuals’ life, instead, they collect location data from Bluetooth, Logger or GPS tracking devices. 
    \item{Unwearable Devices:} Few unwearable devices like personal computers, CCTV are used in lifelogging. Lifelogging from such unwearable devices relies on some attributes like the way of online interaction, emails, image posts, mouse movement, browser history \emph{etc.}  
\end{itemize}
Apart from these sensors which may be required to gather lifelog data, for lifelog processing, some softwares could also be necessary. As lifelogging is a completely multimodal process, for multimodal data processing and training some machine-learning applications, such as Neural Network, might be needed. To run recently trending Neural Network models like Deep or Convolutional Neural Network, for learning the latent representations of multimodal lifelog, GPUs might be required. Alternatively, GPUs could be replaced by cloud computing services like Google Cloud, Amazon Web Services, \emph{etc.} which can ease off the computational complexity or reduce computational time. In case, a pre-trained model for representations is available such a requirement is not necessary. The computers should be equipped sufficiently to avoid long latencies associated with comparing and computing similarities between representations. If Python is used as a programming language, libraries such as tensorflow, numpy, scipy, scikitLearn, pickle might be required.

\subsection{Dataset}
\begin{figure}[ht]
  \includegraphics[width=\textwidth]{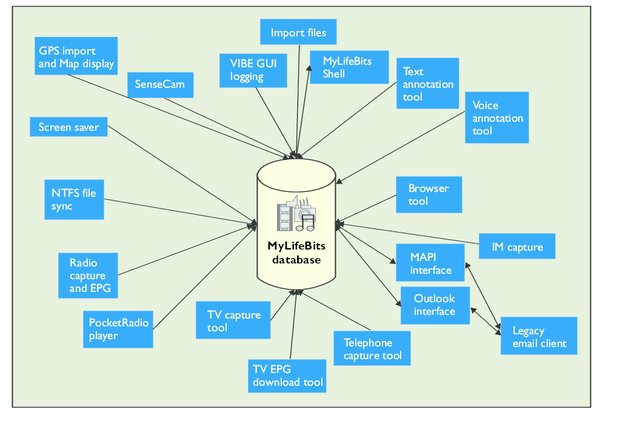}
  \caption{Major Components of MyLifeBits Database}
  \label{fig:MLB}
\end{figure}

A prime requirement towards successful implementation of the project hinges on the data over which the experiment is carried out. While the data may be collected and aggregated from various sources, it should be properly structured and formatted and must also adhere to the need for experiments. We might consider the \textbf{MyLifeBits} data\footnote{https://www.microsoft.com/en-us/research/project/mylifebits/} of Microsoft as a source of lifelog. MyLifeBits which is called `A Personal Database for Everything' is a system that began in $2001$ to explore the use of SQL to store all personal information found in PCs \cite{Gemmell06}. At the very beginning of the project, the system was used to capture and store the scanned and encoded archival material \emph{e.g.} articles, books, music, photos, and video as well as everything born digital \emph{e.g.} office documents, email, digital photos. Gradually, the system unfolded to store everything that could be captured. Thus later, it included web pages, phone calls, meetings, room conversations, keystrokes and mouse clicks for every active screen or document, and around $1-2$ thousand photos, captured by SenseCam every day. Finally in $2006$ the project started to collect real-time data as well as it started using advanced scanning technologies particularly for some sensitive applications like health and wellness.
\paragraph{MyLifeBits Tools:}
As the MyLifeBits database contains everything, there was a need for useful tools to organize, associate metadata, access and report about the information. So, many different capturing and display tools are also associated with the project to populate the store and to aid the search or access it. The main components of MyLifeBits database are depicted in Figure \ref{fig:MLB} \footnote{This figure was uploaded by Researcher Emeritus of Microsoft, Chester Gordon Bell; Source: https://www.researchgate.net/publication/306960201\_MyLifeBits\_A\_personal\_database\_for\_everything/figures}.
Although MyLifeBits dataset is a very competent one but it might not be sufficient for our entire task. Also we might require some recent data. There is another dataset, \textbf{ImageCLEFlifelog}\footnote{Dataset available at http://imageclef-lifelog.computing.dcu.ie/2017/} available which we may consider. \\
The Lifelog dataset consists of data from three lifeloggers for a period of about one month each. The dataset includes $88124$ wearable camera images (Captured frquency: approximately two images per minute), an XML description of $130$ associated semantic locations (\emph{e.g.} Starbucks cafe, McDonalds restaurant, home, work) and few physical activities like walking, cycling, running and transport of the lifeloggers at a granularity of one minute. A brief summary of the content of the dataset is presented in Table \ref{tab:TB}.
\begin{table}
\begin{center}
\begin{tabular}{|c|c|}
\hline
\textbf{Item} & \textbf{Number} \\ \hline  \hline 
Number of Lifeloggers & $3$ \\ \hline
Size of the Collection & $18.18 GB$ \\ \hline
Number of Images  &    $88,124$ \\ \hline
Number of Locations  &    $130$ \\ \hline
Number of Visual Concepts &    $1000$ \\ \hline 
Number of Lifelog Retrieval Task (LRT) Topics &    Development set: $16$, Test set: $20$ \\ \hline
Number of Lifelog Summarization Task (LST) Topics &    Development set: $5$, Test set: $10$ \\ \hline \hline 
\end{tabular}
\end{center}
\caption{Statistics of Lifelog Dataset}
\label{tab:TB}
\end{table}

Given the fact that lifelog data is typically visual in nature and in order to reduce the barriers-to-participation, the output of the CAFFE CNN-based visual concept detector is included in the dataset as additional metadata. This classifier provided labels and probabilities of occurrence for $1000$ objects in every image. The accuracy of the CAFFE visual concept detector is variable and is representative of the current generation of off-the-shelf visual analytics tools. \\
Such datasets may not contain all kinds of data we require. In case, a dataset needs to be prepared according to our needs. It is to be noted that the datasets we are mentioning here are ad-hoc datasets collected over a period and available in a shelf-ready mode. Also, since we plan to incorporate Natural language query processing in several languages, it is essential that such a comparable dataset be available or we may need to augment it with additional data either manually or through crowdsourcing \cite{Li2017}. If the proposed system is supposed to handle real-time data coming from the web or social media, the system needs to be configured accordingly. This alone is a part that can be designated as a separate subtask under the project.
\section{Work Plan}\label{WP}
\begin{itemize}
\item[I.] \textbf{Preparing a dataset}: Once we consider Cross-Lingual Retrieval as a part of our project, we shall require appropriate dataset which exclusively contain Native Language Information. But gathering Native Language Query based lifelog retrieval dataset for which corresponding translated text query is available could be a serious challenge. One way to solve this is to opt for crowd-sourcing or use a translation tool (such as Google Translate API).
\item[II.] \textbf{Generating indexable unit from lifelog}:
Generating indexable unit from lifelogging data is an active research area. Activity-based measures are important and widely used to find the indexable unitfrom lifelog. For instance, in the work by Zhen \emph{et al.}\cite{Zhen13}, a fused data obtained from multiple sensors is used as daily life event segmentation for lifestyle evaluation. Multi-sensor data, motion sensor signals and extracted visual features from images are combined first and then summarized to segment the large  datasets  automatically. Mental and physical activities of an individual is used as a document segmentation model to promote health and wellness in the paper \cite{Braber16}. Mainly, personal behaviour, biometric data such as the food consumption record of an individual, environment records are considered as activities here. This implies that activity-based segmentation is the basic building block of lifelog indexing and it leads us towards better lifelog search and retrieval. \\
Taking cues from the above-mentioned fact, we can define few units which may aid lifelog indexing and retrieval, as follows:
\begin{itemize}
    \item\textbf{Item:} Item is the smallest unit of dataset which can be retrieved. \emph{Ex.} Image, Location, Blood Pressure \emph{etc.}
    \item\textbf{Moment:} Moment is defined as a precise point in time, better saying, a fixed length temporal unit. It can be either minute, hour or any predefined time-slot.
    \item\textbf{Activity:} Activity can be defined as an un-interrupted sequential performance of an individual who is performing a specific function. Activity can be treated as the indexing-time unit of retrieval as though out the activity, an individual is performing a certain task. For example, if we consider 'playing guitar' is an activity, throughout the time the individual is performing the activity, the combination of few items (\emph{Ex.} scene or frame) are taking place repetitively. \item\textbf{Event:} Event can be treated as the longest temporal unit of lifelog and it is the combination of moments or activities. Like, the entire `Musical Evening,' the individual is spending, can be identified as an Event. 
\end{itemize}
These two tasks should not take more than three months of our project time in case the dataset is readily available.
\item[IV.] \textbf{Indoor lifelog data classification}: Before the indexing of items can begin, it would be necessary to classify the recorded visual and non-visual lifelog data into indoor and outdoor events. To this end, low-level image and video features could be employed in order to classify and categorize the videos and images from capturing devices. We plan to draw inspiration from the works of Ledwich and Williams \cite{Ledwich2004} and Lew \cite{Lew2013}. It should be noted that we aim to go beyond the visual cues and posit to categorize other sources of information as well such as geo-location \cite{SUN2017}\cite{Luo18}, time \cite{Wang2017} \emph{etc.}
\item[IV.] \textbf{Query Generation and Reformulation}: Automatically generation of a query from the words present in the topic, would be the main stepping stone. Creation of a minute-by-minute annotation of the individuals' activities with respect to the five identified topics, \emph{namely, diet, exercise, social, where and compare} would serve the purpose \cite{Xu17}. Such topic words can be obtained by applying deep-learning approaches for image analytics and then fusing the multimodal sensor data. Few attributes of individuals' activity such as activity occurrence, temporal and spatial patterns, associations among multiple activities, \emph{etc.} also may help identify the topics.\\

These two steps are the pivotal elements of our project and should take the lion's share of the project time and effort. We plan to finish these two tasks within six to seven months. The rest two or three months would be allocated to the final two tasks and also to write the final report and deliverable.
 
\item[V.] \textbf{Designing the front-end}: If the project is to be delivered as a software application, it would need a front-end with a proper user interface where users can interact with the IR system with minimal knowledge about the back-end.

\item[VI.] \textbf{Validation and assessment}: The final project should be validated by third-party assessor or judges or against a gold standard, if available. The system should be robust against changes in the size of the dataset, should effectively handle outliers, must have low latency and should produce the retrieval results with minimum errors. Again, crowd-sourcing could be a potential tool in this cause.
\end{itemize}

%
\section{Dissemination Plan}
The dissemination encompasses the publishing of scientific and technical articles, papers, posters and oral presentations at relevant international conferences, journals, symposiums, and workshops. The primary criterion for selecting an appropriate venue is the relevance to the subject and area under consideration. The disciplines involved could include (but not limited to) multimedia analytics, wearable and ubiquitous computing, HCI, information retrieval, mobile computing, applications of lifelogging in cognitive science and healthcare and wellness. 
Journals and conferences with high-relevance of the related items of the project and those
that have a strong impact on scientific community and society will be an obvious choice. Some indicative venues could be:
\begin{itemize}
\item[i.] Journal of Visual Communication and Image Representation
\item[ii.] Information Retrieval Journal
\item[iii.] Information Processing \& Management Journal
\item[iv.] ACM SIGIR Conference on Human Information Interaction and Retrieval (CHIIR) 
\item[v.] International Conference on Research and Development in Information Retrieval (SIGIR)
\item[vi.] International Conference on Information and Knowledge Management (CIKM)
\item[vii.] European Conference on Information Retrieval (ECIR)
\item[viii.] ACM CHI Conference on Human Factors in Computing Systems
\item[ix.] IEEE International Conference on Data Engineering (ICDE) 
\item[x.] ACM The International Conference on Mobile Systems, Applications, and Services (Mobisys)
\item[xi.] IEEE International Conference on Mobile Data Management (MDM)
\item[xii.] International Conference on Very Large Data Bases (VLDB)
\end{itemize}
\bibliography{refs}
\bibliographystyle{plain}
\end{document}